\documentclass[prx,aps, twocolumn]{revtex4-1}
\usepackage{graphicx}
\usepackage{wasysym}
\usepackage{amsfonts}
\usepackage{bm}
\usepackage{enumerate}
\usepackage{color}
\usepackage[resetlabels]{multibib}
\usepackage{hyperref}

\usepackage{times}
\newcommand{\beq}{\begin{equation}}
\newcommand{\eeq}{\end{equation}}
\newcommand{\bea}{\begin{eqnarray}}
\newcommand{\eea}{\end{eqnarray}}

\mathchardef\nss="711B

\def\ket#1{{|#1\rangle}}
\def\bra#1{{\langle #1 |}}

\def\vecsymb#1{\boldsymbol{#1}}

\def\nss{\mathcal{S}}

\def\be{\begin{eqnarray}}
\def\ee{\end{eqnarray}}

\newlength{\myL}

\begin{document}
\title{Topological `Luttinger' invariants protected by non-symmorphic symmetry in semimetals}
\author{S. A. Parameswaran}
%\email{sidp@uci.edu}
\affiliation{Department of Physics and Astronomy, University of California, Irvine, CA 92697, USA}
\date{\today}
\begin{abstract}
Luttinger's theorem is a fundamental result in the theory of interacting Fermi systems: it states that the volume inside the Fermi surface is left invariant by interactions, if the number of particles is held fixed. Although this is  traditionally justified in terms of analytic properties of Green's functions, it can be viewed as arising from a momentum balance argument that examines the response of the ground state to the insertion of a single flux quantum [M. Oshikawa, {\it Phys. Rev. Lett.} {\bf 84}, 3370 (2000)]. This reveals that the Fermi volume is a topologically protected quantity, whose change requires a phase transition. However, this sheds no light on the stability or lack thereof of interacting {\it semimetals}, that either lack a Fermi surface, or have perfectly compensated electron and hole pockets and hence vanishing {\it net} Fermi volume. Here, I show that semimetallic phases in non-symmorphic crystals  possess additional  topological `Luttinger invariants' that can be nonzero even though the Fermi volume vanishes.
%I show that spinless or spin-rotation-preserving semimetallic phases in non-symmorphic crystals  possess generalized topological `Luttinger invariants' that can be nonzero even in cases where the Fermi sea volume vanishes.
%
% A nonzero Luttinger invariant then forces energy bands to touch, leading to semimetals whose gaplessness is thus rooted in topology; opening a gap without symmetry breaking automatically triggers fractionalization. 
% 
 The existence of these invariants is linked to the inability of non-symmorphic crystals to host band insulating ground states except at special fillings. I exemplify the use of these new invariants by showing that they distinguish various classes of two- and three-dimensional semimetals.
\end{abstract}
\maketitle
\section{Introduction}
{A} well-trodden path to understanding many-electron systems is to exploit proximity to a well-understood, typically free, model, and use this to access various properties of the interacting system through the lens of perturbation theory. There are few general statements made in this limit that remain valid when the interactions are no longer parametrically small. A celebrated exception to this rule is Luttinger's theorem~\cite{LuttingerOriginal}: it continues to impose constraints on the volume contained within the Fermi surface  --- the momentum-space surface that hosts low-energy excitations ---even when the interactions are strong.  Although Luttinger's theorem is usually derived by examining the properties of Green's functions to all orders in perturbation theory, about a decade ago Oshikawa gave it an elegant topological interpretation~\cite{OshikawaLuttinger}: by determining the change in symmetry of a many-electron ground state upon adiabatically inserting a single quantum of gauge flux, and computing the equivalent response of a Fermi liquid, the electron filling, a quantity that is determined by microscopic physics, can be related to the volume of the Fermi surface -- a property of the low-energy effective theory.  It is this linking of  ultraviolet and infrared scales that makes Luttinger's result a useful tool~\cite{SenthilVojtaSachdev,LuttingerExtension} in analyzing correlated quantum matter. 

In recent years, the class of gapless Fermi systems in $d>1$ has grown to encompass semimetals such as graphene~\cite{RevModPhys.81.109} and its analogs in three dimensions~\cite{PhysRevB.83.205101,YoungetalKaneDSM3d,FangNa3BiDirac,Na3BiExpt1,FangCd3As2Dirac,Cd3As2Expt1}, where the Fermi energy is tuned to intersect a nodal surface where a pair of bands touch, either at discrete points or along continuous contours in the Brillouin zone. Understanding the properties of such semimetals and their descendant phases is a major thrust of current research. The electronic dispersion near the nodal surface is typically linear, leading to a vanishing density of states at the Fermi energy, and consequently such systems are fairly robust: they are perturbatively stable against interaction-induced band gaps, and any correlated gapped phases that emerge from these semimetals do so at fairly strong interactions. It is therefore desirable to place additional constraints on these systems, {\it e.g.} ones similar to those imposed by Luttinger's theorem, that remain valid in the strong-coupling regime. However, semimetals have a vanishing Fermi volume, and so Luttinger's theorem is silent on their behavior. 

\begin{figure}
\includegraphics[width=\columnwidth]{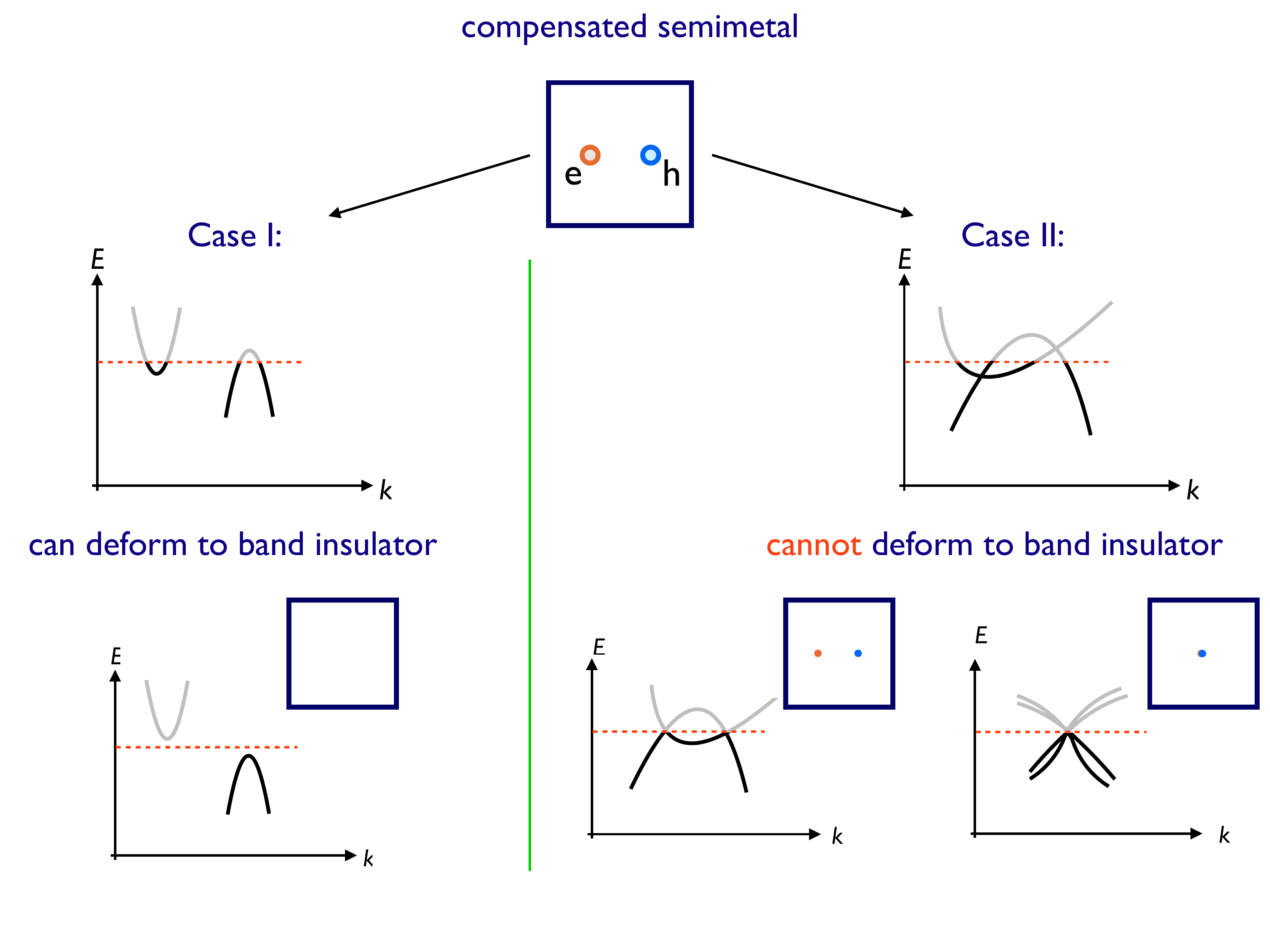}
\caption{\label{fig:deformSM} {\bf Two scenarios for compensated semimetals.} Starting with equal-volume electron and hole pockets it may  be possible to adiabatically deform the system into a band insulator (Case I), or there may be an obstruction to doing so, so that there remain one or more set of nodal points (nodal lines may also be possible) (Case II.) The Fermi surface in each case is shown inset.}
\end{figure}

Two examples serve to illustrate why this is an unsatisfactory state of affairs. First, consider graphene's honeycomb lattice at half-filling. This corresponds to a single electron on each site, and therefore two electrons in each two-site unit cell. Famously, graphene is  a semimetal at half-filling: the Fermi surface consists solely of point nodes, and hence encloses zero Fermi volume. An  band theory analysis reveals that, absent interactions, it is impossible to open gap without breaking symmetry~\cite{HoneycombVoronoi}. However, it {\it is} possible to construct a wavefunction that describes  a gapped, symmetry-preserving phase of {\it interacting} electrons~\cite{HoneycombVoronoi, Ware_TCBI}. This naturally raises the question of whether the existence of such a phase could be inferred on general grounds, rather than by laborious explicit construction --- i.e., whether there is a Luttinger-like theorem for the nodal Fermi surface. A second example  is provided by {\it compensated} semimetals. These are systems whose Fermi surface consists of electron and hole pockets that enclose equal Fermi volumes, so that their {\it net} Fermi volume vanishes. Luttinger's theorem, which is sensitive only to the net Fermi volume, is not readily applicable to this situation, that can arise in one of two ways. The first is to begin with a band insulator, and simultaneously pull the conduction (valence) bands below (above) the Fermi energy, while holding the particle number, and hence the filling, fixed. This `accidental' semimetal can be deformed back into an insulator by reversing this process. Another route begins with a nodal semimetal and pulls nodes above and below the Fermi energy. In this case, the system cannot be adiabatically deformed into a band insulator: its electron and hole pockets can at best be shrunk back to nodal points.  How can we distinguish the two scenarios,  sketched in Fig.~\ref{fig:deformSM},  simply by analyzing the Fermi surface and symmetries of a compensated semimetal?

Here, I show that for certain  semimetals, specifically those that occur in non-symmorphic crystals --- loosely, crystals where screw axes or glide planes are essential to fully characterize the space group --- one can associate additional topological invariants with the Fermi surfaces (either nodal or compensated). I demonstrate their existence via an analysis of crystalline point-group symmetry, focusing on the flow of discrete symmetry charge corresponding to a screw rotation or glide reflection into the system under a flux insertion. The result is an invariant that can be nonzero even when the Fermi sea volume vanishes.  These invariants lead to relations between low-energy parameters and high-energy properties that are analogous to Luttinger's theorem, and hence I dub them `Luttinger invariants'.  Their existence is connected with a higher-dimensional analog of the Lieb-Schultz-Mattis theorem~\cite{Lieb:1961p1,Oshikawa:2000p1,Hastings:2004p1,Hastings:2005p1} that constrains the band structure of non-symmorphic crystals: except at certain specific integer fillings fixed by their space group~\cite{Parameswaran:2013ty,2012arXiv1212.2944R,NonSymSOC}, such crystals cannot become insulating  without either breaking symmetry or else exhibiting {\it topological order}, a phenomenon associated with an emergent gauge structure and the fractionalization of quantum. The flux insertion approach then reveals that the  fractionalized quasiparticles that emerge in the topologically ordered phases transform non-trivially under the crystal symmetry~\cite{FracGlide2D}. These arguments require that the fermions under consideration are spinless, or if spinful, possess an axis of spin conservation;  they do not apply directly to spin-orbit coupled systems that lack this property.

The Luttinger invariants so identified can be used to distinguish between different classes of gapless semimetals. For instance graphene possesses a vanishing Luttinger invariant, that can be computed directly from knowledge of symmetry. This immediately suggests that, with interactions, the semimetal can be gapped. Similarly, compensated semimetals with (without) non-zero Luttinger invariants can (cannot) be deformed into a band insulator without breaking symmetry; we exemplify this dichotomy in $d=2$ in a simple model of spinful electrons on the Shastry-Sutherland lattice, that hosts semimetallic phases both with and without a nonzero Luttinger invariant.

%I exemplify their use in $d=2$ through a simple model of spinful electrons on the Shastry-Sutherland lattice, that hosts semimetallic phases both with and without a nonzero Luttinger invariant. 

%In the case where the invariant is trivial, it is straightforward to show that the semimetallic phase can be gapped out without triggering topological order or breaking the symmetry. In contrast, gapping a semimetal with a nonzero Luttinger invariant  while preserving symmetry inevitably leads to topological order. (The role of the symmetries in the resulting fractionalized phases will be discussed elsewhere~\cite{SungBinMikeSidUnpub}.)
%Similar analyses can be performed for graphene and  gapless semimetals in $d=3$.

The remainder of this article is organized as follows. First, I introduce the flux insertion approach, and show how the non-symmorphic symmetry charge changes upon insertion of a single flux quantum. I illustrate the meaning of this change in terms of the spectral flow of a simple one-dimensional model of free fermions. I then turn to computing this spectral flow in a Fermi liquid, thereby linking the change in symmetry charge to a Fermi surface property. I then demonstrate how knowledge of the Luttinger invariant resolves the puzzles raised in the introduction, before closing with a discussion of applications and extensions.

\section{Flux Insertion and Spectral Flow}
I begin by discussing the case of spinless fermions, before generalizing to spinful electrons (though without spin-orbit coupling). I consider a finite crystal at filling $\nu$ per unit cell, with primitive lattice vectors $\vecsymb{a}_i$ and periodic boundary conditions where $L\vecsymb{a}_i \equiv \vecsymb{0}$, so that the system is defined on a $d$-dimensional torus whose volume is $L^d$ unit cells. Note that here I specialize to the case of integer $\nu$;
as a result $\nu L^d$ is an integer. The situation where $\nu$ is not an integer was discussed previously~\cite{OshikawaLuttinger}. Throughout, I work in units where $\hbar = e =1$, so that the quantum of flux, $\Phi_0 = 2\pi$.

The basic strategy is as follows:  a single quantum of gauge flux is inserted adiabatically into the ground state of the system (alternatively, twisted boundary conditions are imposed around a non-contractible loop on the torus) and then a large gauge transformation is performed to to return the Hamiltonian to its original form. Following Ref.~\cite{OshikawaLuttinger}, I compute how symmetry quantum numbers of the ground state change in the resulting adiabatic cycle using two independent arguments. The first, `trivial symmetry counting', is general and makes no assumptions as to the nature of the ground state --- it depends only on the filling $\nu$. The second will make explicit reference to the ground state, by assuming it to be a Fermi liquid.  Equating the results of these two ways of counting the change in symmetry charge, I arrive at a condition relating the filling to a Fermi surface parameter. It is this condition that depends crucially on the crystal structure. This argument closely parallels that of Oshikawa~\cite{OshikawaLuttinger}, but differs from it in a key respect: rather than the change in momentum of the ground state, here  more general symmetry operators lead to new topologically protected Luttinger invariants
 in the presence of non-symmorphic symmetries.

As a first step, I describe the flux insertion procedure. A necessary and sufficient condition so that the inserted flux is `pure gauge' for any non-contractible loop around the torus is to choose a reciprocal lattice vector $\vecsymb{K} = \sum_{i} m_i \vecsymb{b}_i$, where the $m_i$ are integers, and  $\vecsymb{a}_i\cdot \vecsymb{b}_j = 2\pi \delta_{ij}$, and then pick a gauge in which the Aharonov-Bohm flux $\Phi =  N_\Phi \Phi_0 $ is represented by the uniform vector potential $\vecsymb{A} = N_\Phi \frac{\vecsymb{K}}{L}$. I will assume further that the crystal is invariant under one or more {\it non-symmorphic} symmetry operations $G\equiv \{g|\vecsymb{\tau}\}$ that map points as $G:\vecsymb{r} \rightarrow g\vecsymb{r} + \vecsymb{\tau}$, with $g\vecsymb{\tau} = \vecsymb{\tau}$. Note that a necessary condition for the operation to be non-symmorphic~\cite{Konig:1999p1} is that the translation $\vecsymb{\tau}$ is {\it not} in the lattice of discrete translations, nor is it the projection of any lattice translation into the invariant subspace of $g$. Finally, I will assume that applying $G$ consecutively $\mathcal{S}_G$ times is equivalent to the combination of a point-group operation and a lattice translation, {\it i.e.}, $G^{\mathcal{S}_G}$ is symmorphic; this defines the {\it rank} $\mathcal{S}_G$ of the operation $G$.

\subsection{Counting Symmetry Charge Microscopically}
For each non-symmorphic symmetry $G = \{g|\vecsymb{\tau}\}$, a specific flux insertion that reveals the role of the symmetry can be constructed as follows. Since $G$ always involves a fractional lattice translation $\vecsymb{\tau}$, one can always choose a flux to insert such that the corresponding $\vecsymb{K}$ is the {\it smallest} reciprocal lattice vector parallel to $\vecsymb{\tau}$; note that it follows that $\vecsymb{K}$ is left invariant by $g$, {\it i.e.}, $g\vecsymb{K} = \vecsymb{K}$. 
 Flux is then inserted adiabatically into the system by switching on a time-dependent vector potential $\vecsymb{A}(t)= f(t) \frac{\vecsymb{K}}{L}$, such that $f(0)=0$ and $f(T) =1$; at the end of the time $T$, there will be exactly one quantum of flux enclosed by a loop that encircles the torus parallel to $\vecsymb{K}$. The ground state $\ket{\Psi_0}$ at $t=0$ is assumed to preserve all the space group symmetries. Therefore it has a definite $G$-eigenvalue, given by $\hat{G}\ket{\Psi_0} = e^{iG_0}\ket{\Psi_0}$, where $\hat{G}$ is the unitary operator that implements the symmetry $G$ on the Hilbert space. Since the Hamiltonian commutes with $G$ at any $t \in [0,T]$ in this choice of gauge, it follows that at the end of the adiabatic flux insertion, the system is in some state $\ket{\Psi_0'}$ that has the same $G$-eigenvalue as the ground state, {\it i.e.} $\hat{G} \ket{\Psi_0'} = e^{iG_0}\ket{\Psi_0'}$. The final step is to return to the original gauge, by performing the large gauge transformation
\be\label{eq:gaugetrans}
\hat{U}_{\vecsymb{K}} =  \exp\left[-\frac{i}{L}\int d^d r (\vecsymb{K} \cdot \vecsymb{r})\hat{\rho}(\vecsymb{r}) \right].
\ee
 It is straightforward to demonstrate that 
\be
\hat{U}^{-1}_{\vecsymb{K}} \hat{G}\hat{U}_{\vecsymb{K}} = \hat{G} \exp\left[i\frac{{\vecsymb{\tau}\cdot\vecsymb{K}}}{L}\int d^d r \hat{\rho}(\vecsymb{r})\right]. 
\ee
From this, it follows that the a full cycle of inserting a flux and performing the gauge transformation (\ref{eq:gaugetrans}) yields a state $|{\tilde{\Psi}_0}\rangle = \hat{U}_{\vecsymb{K}} \ket{\Psi_0'}$, an eigenstate of $\hat{G}$ with eigenvalue $e^{i G}$, where
\be\label{eq:trivialcountinggenerictau}
e^{i {G}} = e^{i \left(G_0 + \nu L^{d-1}  \vecsymb{\tau}\cdot\vecsymb{K}\right)}.
\ee
 Eq. (\ref{eq:trivialcountinggenerictau}) 
 gives the {universal} change in crystal symmetry charge upon a flux insertion, independent of how this change is accommodated in the system, {\it i.e.}, independent of the phase of the system. 
 
 \subsection{Spectral Flow: a one-dimensional example}
 
\begin{figure}[tb]
\centering
\includegraphics[width=0.9\columnwidth]{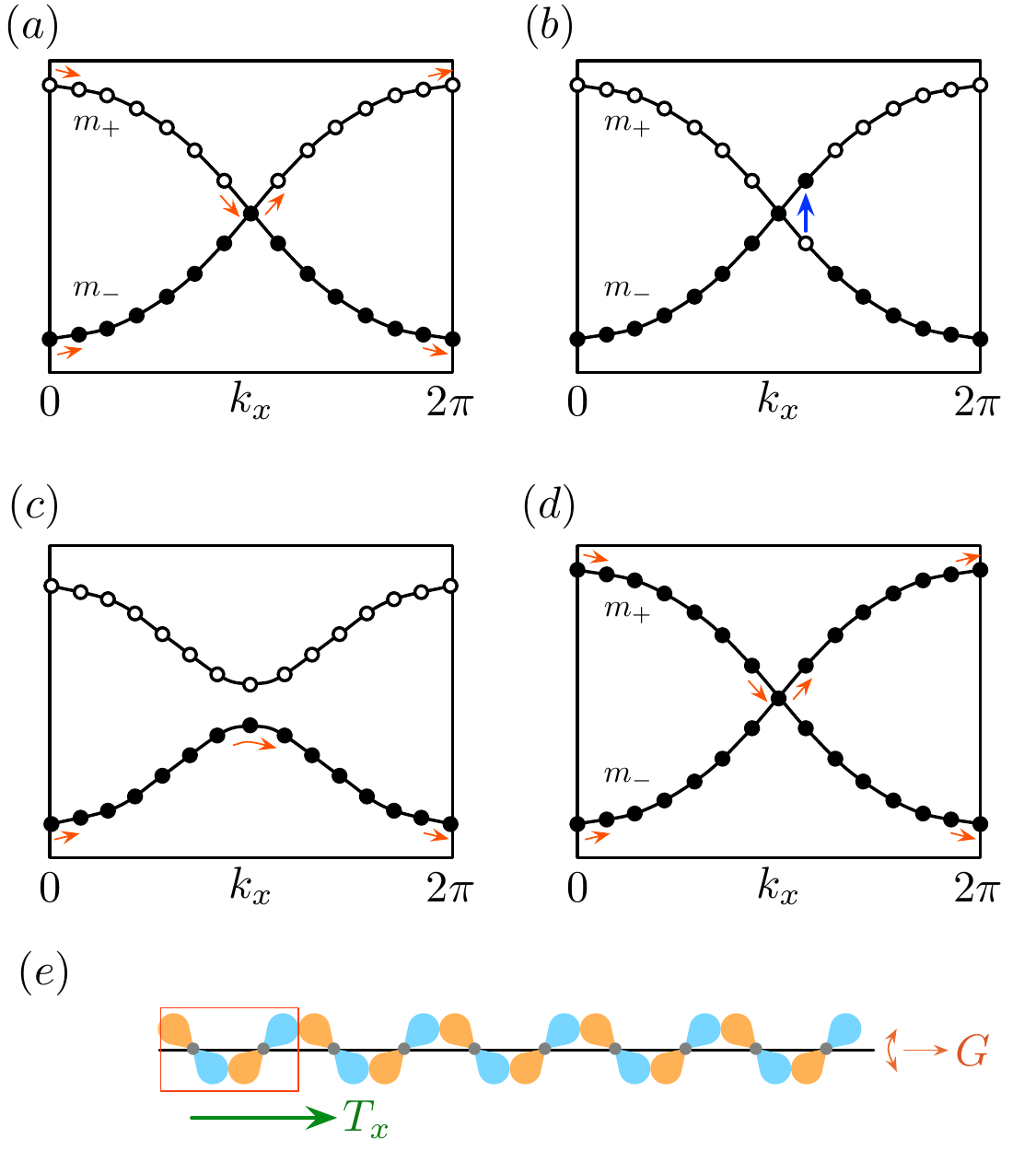}
\caption{{\bf Spectral Flow at Integer Filling in $d=1$.} (Brillouin zone shown in extended-zone scheme) (a) Free fermion dispersion for 1D model with translation $\hat{T}_x$ and a single internal symmetry $\hat{G}$ involving a half-translation. Bands are labeled by their $G$-eigenvalue $m_{\pm 1}$. At $\nu=1$ all states are filled up to the nodal point, which is doubly-degenerate and therefore half-filled; upon flux insertion, filled quasiparticle states flow adiabatically (red arrows.). (b) Occupied states after flux insertion at $\nu=1$: one charge has been transferred from valence to conduction band, leading to a shift in symmetry charge. Flux insertion leads to trivial spectral flow at $\nu=2$ (c) or at $\nu=1$ when the symmetry $G$ is broken and the bands are detached (d). A representative 1D $p$-orbital lattice model with this symmetry is shown in (e) with the unit cell denoted by a red box.} 
\label{fig:fluxins}
\end{figure}

Before proceeding, a simple  example may serve to illuminate the meaning of the Luttinger invariant.
Consider a one-dimensional  lattice with two symmetries: translation ($\hat{T}_x$) and a `non-symmorphic' symmetry $\hat{G}$ that involves a discrete symmetry coupled with a half-translation --- for instance, one could imagine a lattice with a two-site unit cell with reflection-conjugate orbitals on alternating sites (Fig.~\ref{fig:fluxins} (e)). (Strictly speaking, there are no true non-symmorphic symmetries in $d=1$, but this toy model serves to illustrate the general principle.)  In units where the lattice spacing is $a=1$, we have ${\tau} =\frac{1}{2} $, and ${K} = {2\pi}$.  
Now, consider non-interacting spinless fermions at $\nu=1$ per unit cell, so that each site is half-filled. 

A glide mirror must square to a lattice translation: in the present example, $\hat{G}^2 = \hat{T}_x$. The 1D Bloch hamiltonian $h(k)$ must therefore satisfy $\mathcal{M}_G(k)h(k) \mathcal{M}_G^{-1}(k)= h(k)$, where $\mathcal{M}_G(k)$, which represents the action of the glide mirror on the Bloch states, is a function of $k$, a necessary condition for a glide. It is straightforward to see that  $\mathcal{M}_G(k) = \sigma_x e^{i k/2}$, where $\sigma_x$ represents the action on the orbitals, and the $e^{i k/2}$ ensures that $[\mathcal{M}_G(k)]^2 = e^{i k} = T_{x}$ in the Bloch basis. Thus, the Bloch states can be labeled by their eigenvalues $m_{\pm}(k) = \pm e^{i k/2}$. Crucially, upon shifting $k\rightarrow k+2\pi$, the eigenvalues switch places, which requires that the two branches cross an {\it odd} number of times as the Brillouin zone is traversed (in higher dimensions, this would count the number of crossings along the glide plane.) With the further assumption that the model respects inversion symmetry,  the crossing must occur at the zone boundary, $k=\pi$;  on these general grounds, the band structure must take the form shown in Fig.~\ref{fig:fluxins}a (Note that the figure uses a slightly unconventional choice of the Brillouin zone, showing $k \in [0, 2\pi]$ rather than the more familiar choice of $[-\pi, \pi]$, to better illustrate the spectral flow.) At $\nu=1$, the chemical potential lies at the crossing point of the two bands, so that the Fermi surface consists of a single point. 

Now, consider the spectral flow induced by flux insertion, which results in moving each single-particle state via ${k} \rightarrow k+ \frac{2\pi}{L}$. At the nodal point (that we will denote $k_0$ in the general case), $-k$ states from the valence band must evolve into $+k$ states in the conduction band, as flux insertion preserves their $G$-eigenvalue. Similarly, $-k$ states in the conduction band evolve into $+k$ states in the valence band (Fig.~\ref{fig:fluxins}(a)). From this, it follows that the net result of the flux insertion is to transfer a single charge from the conduction to the valence band (Fig.~\ref{fig:fluxins}(b)) at the momentum $k_{\text{exc}}= k_0 + \frac{2\pi}{L}$ (In the thermodynamic limit, $k_{\text{exc}}\rightarrow k_0$.). Thus, after the flux insertion we have $e^{i(G-G_0)} =  m_-(k_{{\text{exc}}})m_+^*(k_{{\text{exc}}}) = -1$, so that $G-G_0 = \pi\,\text{(mod $2\pi$)}$, consistent with the microscopic counting in the preceding section for $\nu=1$. Observe that a similar conclusion obtains as long as the bands cross an odd number of times.

 If it were possible to detach the bands without breaking symmetry (Fig.~\ref{fig:fluxins}(c)), then the corresponding spectral flow would simply involve the fully filled valence band, and there would be no change in the symmetry quantum number of the ground state. The latter possibility is inconsistent with the microscopic counting: in other words, the non-symmorphic symmetry {\it requires} that the bands cross. Note however that at $\nu=2$, where both bands are filled, the adiabatic cycle is trivial and does not change the symmetry quantum number (Fig.~\ref{fig:fluxins}(d)), but this is consistent with the counting from the microscopic theory. Thus, the microscopic counting for $\nu=1$ constrains the band structure to require an odd number of crossings.
 
 This spectral flow picture may be generalized to $d>1$ by considering effective 1D band structures for each distinct transverse momentum.  We note that a similar discussion of the band structures required by non-symmorphic symmetries is given in Ref.~\cite{YoungKaneDSM2d}, albeit without the discussion of spectral flow here. Ref.~\cite{PhysRevB.91.161105} discusses a closely related spectral flow on the surface of a topological crystalline insulator where one of the protecting symmetries is a glide mirror. 

An alternative way to phrase the above discussion is to simply require that the the  change in crystal symmetry quantum number computed in terms of the shift of quasiparticle states under $k\rightarrow k + \frac{2\pi}{L}$ agrees with the microscopic counting. If the microscopic counting involves a nontrivial symmetry change, then it follows (by arguments similar to the example above) that this must be reflected in the low-energy description. If the system is in a Fermi liquid phase, this gives an invariant associated with the Fermi surface response to flux insertion. I now adopt this approach to compute a `Luttinger invariant' associated with a Fermi surface of zero volume for $d\geq 1$.
 
 \subsection{Counting Symmetry Charge in the Fermi Liquid}
I now compute the flow of symmetry charge upon flux insertion {assuming} the system is in a Fermi liquid phase, and so can be described in terms of long-lived quasiparticles, with low-energy effective Hamiltonian
 \be\label{eq:FLHam}
 \mathcal{H}\sim \sum_{\vecsymb{k}}\epsilon(\vecsymb{k}) \tilde{n}_{\vecsymb{k}} + \sum_{\vecsymb{k},\vecsymb{k}'} f(\vecsymb{k},\vecsymb{k}')\tilde{n}_{\vecsymb{k}}\tilde{n}_{\vecsymb{k'}},
 \ee
Assume that the ground state before flux insertion is the filled Fermi sea of quasiparticle states. Under the flux insertion, any change in the ground state can be accounted for in terms of quasiparticle excitations generated near the Fermi surface. Recall that quasiparticles near the Fermi surface are long-lived; using this along with the adiabatic continuity with the Fermi gas, one may determine the change in the quasiparticle population $\delta\tilde{n}_{\vecsymb{k}}$. Observe that flux threading for noninteracting fermions simply shifts the Fermi sea by a uniform amount $\frac{\vecsymb{K}}{L}$, as each $\vecsymb{k} \rightarrow \vecsymb{k}  + \frac{\vecsymb{K}}{L}$. Now, a uniform shift of the Fermi surface of this form amounts to producing quasiparticles on one side of the Fermi surface, and quasiholes on the opposite side. Therefore, as the excitations are close to the Fermi surface,  Fermi liquid theory may be used to compute the change in the ground-state properties.

For this purpose, it is convenient to split the behavior of the glide operation into its two distinct components: the point group operation and the fractional translation, by writing $\hat{G} = \hat{g}\hat{T}_{\vecsymb{\tau}}$. Consider the states $\ket{\Psi_0},\ket{\tilde{\Psi}_0}$ before and after flux insertion. Then,  (using similar notation to the previous section)
\be
e^{i G_0} &=& \bra{\Psi_0}\hat{g}\hat{T}_{\vecsymb{\tau}} \ket{\Psi_0}, e^{i G}=  \bra{\tilde{\Psi}_0}\hat{g}\hat{T}_{\vecsymb{\tau}} \ket{\tilde{\Psi}_0}.
\ee
Now, observe that the shift of the Fermi surface was by a $g$-invariant momentum (recall $g\vecsymb{K} = \vecsymb{K}$), so it follows that the only difference between $G$ and $G_0$ emerges from the fractional translation by $\vecsymb{\tau}$. Since  the result of the adiabatic process is the shift of the whole Fermi sea by $\vecsymb{K}/L$, the change in the momentum $\vecsymb{P}$ of the system during the adiabatic process is given by~\cite{OshikawaLuttinger,SenthilVojtaSachdev, LuttingerExtension}
\be\label{eq:momentumchange}
\Delta\vecsymb{P} = N_F^{(L)}  \frac{\vecsymb{K}}{L}
\ee
where $N_F^{(L)}$ is the integer `occupation' number of filled quasiparticle states in the ground state. (I note that a more precise computation actually obtains $\Delta\vecsymb{P}$ by first summing over deformations of the quasiparticle distribution near the Fermi energy, converting the sum to an integral over the Fermi surface, and then using Stoke's theorem to relate this to the Fermi surface volume, and hence the number of filled quasiparticle states relative to the bottom of the conduction band~\cite{OshikawaLuttinger, SenthilVojtaSachdev, LuttingerExtension}. As this yields the same result for $L\rightarrow\infty$, I simply use the simple formula above.) From (\ref{eq:momentumchange}) and the preceding discussion it is possible to compute the change in symmetry charge in a Fermi liquid:
\be\label{eq:FLmomentumcounting}
e^{iG} = e^{i\left(G_0 + {N^{(L)}_F}\frac{\vecsymb{\tau}\cdot\vecsymb{K}}{L}\right)}.
\ee
where we have used the fact that the fractional translation has eigenvalues $e^{i \vecsymb{\tau}\cdot\vecsymb{P}}$. An explicit calculation of the symmetry charge in the free fermion case is provided in an appendix.%Appendix~\ref{app:FFGsym}.

Eq.~(\ref{eq:FLmomentumcounting}) expresses the change in the symmetry charge assuming that the low-lying excitations form a Fermi liquid. Comparing this to the microscopic counting in (\ref{eq:trivialcountinggenerictau}), %we obtain a
yields a consistency condition, namely that
$
 ({N_F^{(L)}}{L}^{-1} - \nu L^{d-1}) \vecsymb{\tau}~\cdot~\vecsymb{K} 
$ is an integer multiple of $2\pi$.

To proceed, %we need to
one must  determine the value of $\vecsymb{\tau}~\cdot~\vecsymb{K}$. This depends crucially on the fact that $G$ is non-symmorphic.
From the definition of the rank, $G^{\mathcal{S}_G}$ is symmorphic, so that the associated translation $\mathcal{S}_G\vecsymb{\tau}$ must be a lattice translation. Since $\vecsymb{K}$ is the smallest reciprocal lattice vector parallel to $\hat{\vecsymb{K}}$, it follows at
that  $\vecsymb{\tau}\cdot\vecsymb{K} = 2\pi p/\mathcal{S}_G$ for some integer $p<\mathcal{S}_G$, relatively prime to $\mathcal{S}_G$. 
From this, %we find that
\be
p ( N_F^{(L)} - \nu L^{d}) =  L\times (\text{integer}) \times \mathcal{S}_G.
\ee
Let us take $L$ relatively prime to $\mathcal{S}_G$, and define $\chi_F^{(L)} =  N_F^{(L)} L^{-d}$, so that 
\be \label{eq:asymL}
p (\chi_F^{(L)} - \nu ) L^d=  L\times (\text{integer})\times \mathcal{S}_G.
\ee
Since the number of filled quasiparticle states in the system is extensive, it follows that as $L\rightarrow\infty$, $\chi_F^{(L)}$ is independent of $L$. Since $\nu$ is also independent of $L$, (\ref{eq:asymL}) cannot be consistent in the thermodynamic limit unless the RHS also scales with~\footnote{The exception is if $\chi_F^{(L\rightarrow\infty)}=\nu$ in which case both sides vanish as $L\rightarrow\infty$, but this just yields a special case of the more general result.} $L^d$. Thus, 
canceling a factor of $L^d$  from both sides of (\ref{eq:asymL}) leads to  
 $p (\chi_F^{(L)} - \nu ) =  (\text{integer})\times \mathcal{S}_G$.
 Finally, as $p$ and $\mathcal{S}_g$ are relatively prime, the only way to satisfy this relation is if
\be\label{eq:mainresultmod}
\chi_F = \nu - n\mathcal{S}_G
\ee
for $n$ any integer (I have dropped the superscript on $\chi_F^{(L)}$ with the understanding that I refer to its $L\rightarrow\infty$  limit.)

\section{Luttinger Invariant}
In a crystal whose space group contains many non-symmorphic symmetries, one may obtain a similar constraint on $\chi_F$ for each such symmetry. From this and the fact that $n$ can be any integer, it follows that if $\mathcal{S}^*$ is the 
the least common multiple of all the $\mathcal{S}_G$, then
\be\label{eq:mainresult}
\tilde{\chi}_F =  \nu\, (\text{mod } \mathcal{S}^*)
\ee
is a topological invariant of the system, that I propose to term a `Luttinger invariant'. For any crystal containing non-symmorphic operations, $\mathcal{S}^*>1$; of the 157 non-symmorphic space groups, 155 are of this type. However, there exist two `exceptional' non-symmorphic space groups where every individual operation can be rendered symmorphic by a change in real-space origin, so that the argument above does not apply immediately. Still, for these groups it is possible to show suitable {\it combinations} of consecutive flux insertions yield similar constraints, so that they too have $\mathcal{S}^*>1$ (both have $\mathcal{S}^*=2$). Therefore, $\mathcal{S}^*>1$ if and only if the 
crystal is non-symmorphic.

Eq. (\ref{eq:mainresult}) (or equivalently, (\ref{eq:mainresultmod})) is the central result of this paper: it relates the filling, defined in the microscopic theory, to the counting of quasiparticle excitations in the emergent low-energy description (\ref{eq:FLHam}). The second term on the right-hand side of (\ref{eq:mainresultmod}) refers to the filled bands; in essence, this equation reflects the fact that all the charge in the system is either bound up into filled bands or contributes to the low-energy gapless quasiparticle excitations. The presence of $\mathcal{S}^*$ in this expression is linked to a topological requirement that energy bands in crystals can only appear in multiplets (containing a multiple of $\mathcal{S}^*$ bands) that `stick' together~\cite{Parameswaran:2013ty}. 
Note that there is a subtle distinction between the {\it non-symmorphic rank} $\mathcal{S}$ of a space group as defined in Ref.~\cite{Parameswaran:2013ty} and the definition of $\mathcal{S}^*$ used here; I comment on this in the appendix, %Appendix~\ref{app:rankcomments}, 
but this distinction is unimportant for the examples studied here.

An expression similar to Eq. (\ref{eq:mainresult}) was derived in Ref.~\cite{OshikawaLuttinger}: if $V_F \equiv (2\pi)^d\chi_F$, denotes the volume of the Fermi sea, then $\frac{V_F}{(2\pi)^d} -\nu$ must be an integer in a translationally-invariant system. For fractional $\nu$, this is the familiar result that the Fermi sea volume is protected: it is proportional to the fractional part of the filling. When the filling is an integer (say $\nu=1$, for specificity) the {volume} of the Fermi sea vanishes. This is in accord with the intuition (to be qualified shortly) that a system with all states in the Brillouin zone filled is inert and can be adiabatically deformed into one with no gapless excitations and hence no Fermi surface. In this manner,  Luttinger's theorem can be seen to have a topological origin: the volume of the Fermi sea is an invariant that is `protected' against interactions as long as the low-energy description takes the form of Eq.~(\ref{eq:FLHam}). Extending this picture to include more general symmetries beyond translation, it is clear that $\mathcal{S}^*=1$, corresponding to a symmorphic crystal, is uninteresting: there are no additional Luttinger invariants besides the Fermi sea volume. When this vanishes, there is  indeed no obstruction to adiabatically deforming the system into a gapped phase.  

Non-symmorphic crystals where $\mathcal{S}^*>1$ possess additional invariants that modify this picture. From  (\ref{eq:mainresult}) one can see that as long as $\nu$ is indivisible by $\mathcal{S}^*$, it must be true that $\tilde{\chi}_F = \nu \,(\text{mod } \mathcal{S}^*) \neq 0$. As we have noted, for any integer $\nu$, the volume of the Fermi sea  vanishes in the Brillouin zone, so $\tilde{\chi}_F$ must be a new invariant {\it distinct} from the Fermi sea volume. Observe that $\tilde{\chi}_F$ counts the symmetry change induced by low-energy quasiparticles described by (\ref{eq:FLHam}); thus, the fact that $\tilde{\chi}_F$ is  non-zero means that these excitations must be gapless. (For, if they were gapped, it would be possible to smoothly change parameters until they were very far away from the Fermi surface and therefore could not contribute to $\tilde{\chi}_F$.) One way to reconcile the vanishing of the Fermi sea volume with  the gaplessness of (\ref{eq:FLHam}) is if the Fermi surface consists entirely of point or line nodes, {\it i.e.}, describes a nodal semimetal. An alternative is a compensated system, with equal-volume electron and hole Fermi surfaces, although this may be ruled out by additional symmetries~\footnote{Any symmetry that interchanges the location of the disconnected components without charge conjugation forbids this.}. Rather than the momentum balance that leads to the protection of a Fermi sea, here the topological protection is linked to the quantum numbers of discrete spatial symmetries such as rotation or reflection.
Thus, %the presence of 
non-symmorphic symmetries allow %s
 us to identify a class of %nodal 
 semimetals that are protected by a topological `Luttinger' invariant analogously to how a filled Fermi sea is protected by Luttinger's theorem. 
 
 Note that, as in the case with a nonzero Fermi sea volume, one way to evade the protection is that the low-energy theory no longer takes the form of a Fermi liquid so that  (\ref{eq:FLHam}) is no longer a valid description of the system. However, even a non-Fermi liquid without sharply-defined quasiparticles, that nevertheless has an appropriately defined Fermi surface of low-energy excitations, is still amenable to this argument and therefore can have a nonzero Luttinger invariant; similar results then apply. There is a familiar example in $d=1$: Luttinger liquids  satisfy Luttinger's theorem, but are not Fermi liquids~\cite{PhysRevLett.79.1106,PhysRevLett.79.1110}. Alternatively, the system could open a gap, but cannot enter a trivial insulating phase (meaning one whose wave function can be adiabatically continued to that of a band insulator) without breaking symmetry~\cite{Parameswaran:2013ty}, reflecting the generalized Lieb-Schultz-Mattis theorem. Note that all these arguments require the presence of $U(1)$ charge conservation and the space group symmetries; however, the breaking of these symmetries is usually detectable in experiments.

This picture shows  the crucial role played by the non-symmorphic symmetry: on the one hand, it has a nontrivial relationship with translation, meaning that adiabatic shifts of momenta change its value even for integer $\nu$; on the other hand, it is defined along a line or plane in the Brillouin zone, and therefore we can track the spectral flow of quasiparticles in a concrete way by assigning definite symmetry quantum numbers to bands along these high-symmetry directions.  Note that the analysis of band structures is limited to non-interacting systems, but the arguments that led to the computation of the Luttinger invariant are non-perturbative in nature, and so apply more generally. The non-perturbative Luttinger invariant (\ref{eq:mainresult}) provides a simple way to compute, purely from knowledge of stoichiometry and crystal structure, whether a given semimetallic dispersion must exhibit a nontrivial spectral flow even in the presence of interactions, and hence identify protected semimetals.

\subsection{Including Spin}
The extension of these arguments to electrons with spin is straightforward in the absence of spin-orbit coupling. Since the two spin species are independently conserved, one may define separate fillings so that $\nu = \nu_{\uparrow} + \nu_{\downarrow}$. Accordingly, it is possible to introduce a fictitious gauge flux that couples to the up and down spins separately, and follow the reasoning above to obtain a pair of invariants, $\tilde{\chi}^\sigma_F = \nu_\sigma\,(\text{mod } \mathcal{S}^*)$ with $\sigma= \uparrow, \downarrow$. In the spin symmetric case where $\nu_\uparrow =\nu_\downarrow$, we may simply write $\tilde{\chi}_F  = \tilde{\chi}^\uparrow_F = \tilde{\chi}^\downarrow_F = \frac{1}{2} \nu\,( \text{mod } \mathcal{S}^*)$: even though there are twice as many electrons in each unit cell because of the spin degeneracy, in this case the topological invariant is computed by simply halving the filling and therefore coincides with that computed for spinless fermions. Note that, in applying this argument, we assume that the spin transforms trivially under the mirror component of the glide symmetry --- a choice that is only consistent absent spin-orbit coupling, that breaks the spin rotation symmetry. 

\section{Examples}

We now revisit the examples from the introduction, armed with the new Luttinger invariant.
First, consider graphene, where the spin-orbit coupling is negligible and can be set to zero. At half-filling on the honeycomb lattice, $\nu=\nu_\uparrow+\nu_\downarrow =2$, owing to the two sites in each unit cell; therefore,  the Fermi volume vanishes. As the honeycomb lattice has a symmorphic space group ($P6/mmc$), it has no other nontrivial Luttinger invariants; and hence no generalized `Luttinger theorem' can be associated with Dirac nodes in graphene. This is consistent with the existence of a gapped, symmetry-preserving phase of spinful fermions at $\nu=2$ on the honeycomb lattice with no fractionalization~\cite{HoneycombVoronoi, Ware_TCBI}; such a phase cannot descend from a gapless system with a nonzero Luttinger invariant. Thus, knowledge that the Luttinger invariant vanishes allows one to deduce the existence of a gapped symmetry-preserving phase without explicit construction of a microscopic model or trial wavefunction.

\begin{figure}[tb]
\centering
\includegraphics[width=\columnwidth]{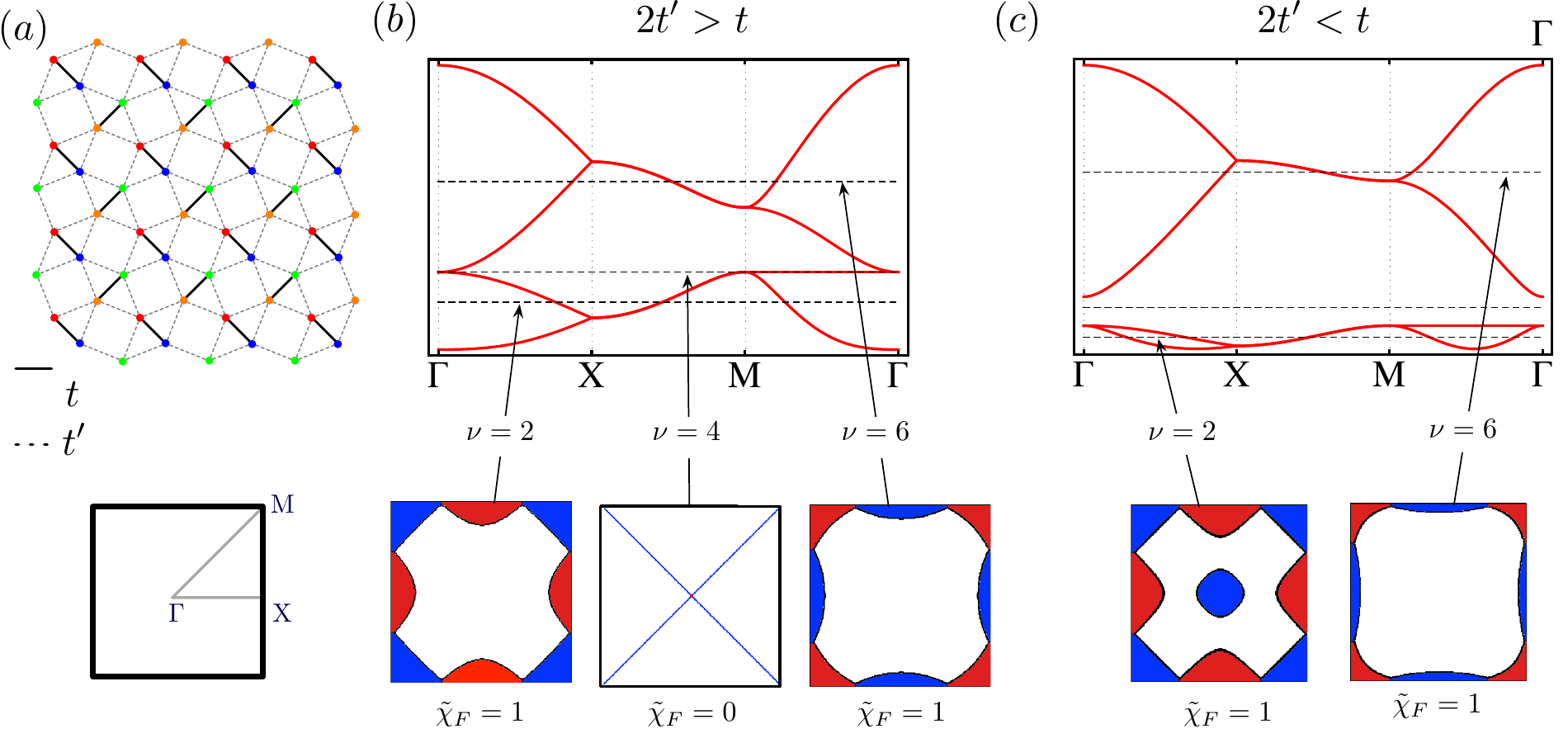}
\caption{{\bf Luttinger Invariants in the Shastry-Sutherland Lattice}. (a) Shastry-Sutherland lattice with $p4g$ space group and Brillouin zone with high-symmetry directions labeled. (b,c): Tight-binding band structure and Fermi surfaces at even integer fillings for representative values of $t'/t$; electron (hole) pockets are colored in red (blue). Owing to the combination of inversion  and time-reversal symmetry, each energy level is two-fold degenerate.
For $2t'>t$, (b) the system is a (perfectly compensated) semimetal at one-quarter-, half- and three-quarter- filling ($\nu=2,4,6$ for spinful electrons.). For $\nu=2,6$ the Fermi surface consists of electron and hole pockets enclosing zero net area, whereas $\nu=4$ it consists of a quadratic band crossing at the zone center and a non-dispersing line of filled electronic states connecting the zone center to the corners. The semimetals at $\nu=2,6$ semimetals are protected by a nonzero Luttinger invariant ($\tilde{\chi}_F =1$) and remain gapless as we tune parameters without changing symmetry, but the unprotected nodes can be gapped, as showj in (c), where $2t'<t$. In the strongly interacting limit at half-filling the effective description is a Heisenberg model whose (unique) ground state is a crystal of singlets on strong bonds~\cite{ShastrySutherland} (solid lines in (a)); this phase can be adiabatically connected to the $\nu=4$ band insulator in (c).  the non-zero Luttinger invariant guarantees that any gapped symmetric ground state at $\nu=2,6$ has topological order. Note that the fact that the electron and hole pockets themselves touch in the Brillouin zone is an artefact of the neglect of spin-orbit and higher-neighbor couplings.} 
\label{fig:SSL}
\end{figure}

Turning to our second puzzle of compensated semimetals, it is useful to focus on an illustrative example in $d=2$ provided by electrons on the Shastry-Sutherland lattice (SSL)~\cite{ShastrySutherland}, again without spin-orbit coupling. The SSL has a non-symmorphic space group ($p4g$) with $\mathcal{S}^* =\mathcal{S} =2$. Using a simple $s$-orbital tight-binding model we find that the non-interacting band structure may exhibit nodal semimetallic behavior at even integer values of $\nu = \nu_{\uparrow} + \nu_{\downarrow}$. Consider the half-filled case, where  $\nu=4$, so that  $\nu_\uparrow =\nu_\downarrow = 2$ (Fig.~\ref{fig:SSL}(b)). From the preceding arguments, at this filling there is no Luttinger invariant associated with the energy bands: $\tilde{\chi}_F= \frac{1}{2} \nu\,( \text{mod } \mathcal{S}^*) = 0$. Indeed, it is straightforward to write down a simple insulating wavefunction at this filling~\cite{ShastrySutherland}, by binding electrons into singlets placed on the solid bonds in Fig.~\ref{fig:SSL}(a); alternatively, suitable choices of the hopping parameters leads to a gap opening (Fig.~\ref{fig:SSL}(c)) without a symmetry change. On the other hand, at quarter filling ($\nu =2$, $\nu_\uparrow =\nu_\downarrow = 1$) the considerations presented here reveal that the semimetallic phases are protected by a non-zero Luttinger invariant ($\tilde{\chi}_F = 1$), and therefore cannot be gapped without triggering fractionalization or breaking symmetry. While semimetallic phases 
that appear in the SSL have been studied recently~\cite{HatsugaiSSL}, the essential distinction between those at $\nu=2$ and $\nu=4$ seems to have been overlooked. 

Similar statements can be made for three-dimensional non-symmorphic crystals, {\it e.g.} the diamond or hexagonal close-packed structures. Specifically, any {\it filling-enforced semimetal}~\cite{FESM} in a non-symmorphic crystal  is associated with a non-zero Luttinger invariant. To see this, we observe that a filling-enforced semimetal occurs in a non-symmorphic crystal at any filling (of spinful electrons) that is not an even integer multiple of the non-symmorphic rank; the non-zero value of the Luttinger invariant then follows immediately from the arguments of the preceding section. Technically, our identification of the Luttinger invariant relies on the presence of spin-rotation symmetry and hence the absence of spin-orbit coupling, that may be present for a generic filling-enforced semimetal. However, motivated by the absence of symmetry-preserving, non-fractionalized insulating ground states at such fillings even with spin-orbit interactions~\cite{NonSymSOC}, I conjecture that  the Luttinger invariant remains non-zero also in this case.

\section{Concluding Remarks}
In closing, I comment briefly on  the applicability of these ideas to three dimensional nodal semimetallic phases. Recently,  Dirac semimetallic phases have been identified in non-symmorphic crystals in two~\cite{YoungKaneDSM2d} 
and three~\cite{YoungetalKaneDSM3d,PhysRevLett.112.036403, FESM} dimensions. The corresponding band structures would have non-zero Luttinger invariants in the limit of vanishing spin-orbit coupling; however, in this limit the bands can touch along nodal surfaces rather than at isolated points, and it is not clear whether the invariants as defined here survive the inclusion of spin-orbit coupling.  Note that a quite different approach~\cite{NonSymSOC} than flux-threading seems necessary to extend similar arguments for gapped phases to systems without spin-rotation symmetry; whether such techniques can be suitably adapted to treat gapless systems remains an outstanding problem. Therefore, whether a similar invariant can be identified in the presence of strong spin-orbit coupling is at present an open question, that seems worthy of further study.

 It is possible to extend similar ideas to Kondo lattice models believed to capture the essential physics of heavy-fermion materials. Here, a generalized Luttinger invariant can be computed if the filling is taken to count both the conduction electrons and the local spin moments within the unit cell. The derivation of this invariant and its consequences for heavy fermion systems are beyond the scope of the present article, but will be discussed elsewhere~\cite{2016arXiv160904023P,BPLP-unpub}.
 
Finally, I emphasize that while Dirac semimetals with nodes on their zone boundaries do not generically have topological surface states or a nontrivial bulk electromagnetic response --- and are thus not `topological' in one sense that is currently in vogue --- the presence of a nonzero Luttinger invariant means that they are {parent semimetals} for three-dimensional {\it topologically ordered} ({\it i.e.}, fractionalized) phases that emerge when they are gapped while preserving symmetry. This suggests that understanding such semimetals and their instabilities is one possible route to new phases of matter. 
\newline\\
\noindent{\bf Acknowledgements.} I thank SungBin Lee, Michael Hermele, Ari Turner, Daniel Arovas and Ashvin Vishwanath for discussions and collaboration on related work, and Ashvin Vishwanath and especially Mike Zaletel for illuminating discussions and comments on the manuscript. I acknowledge support from the National Science Foundation via Grant No. DMR-1455366, and UC Irvine start-up funds.

\begin{appendix}

\section{\label{app:FFGsym} Counting Symmetry Charge for Free Fermions}

Although the argument in the main body of the paper that restricted consideration to the low-energy excitations near the Fermi surface (where there are well-defined, scattering free quasiparticles), the conclusion was that the change in symmetry charge upon flux insertion was simply related to the number of {\it filled} momentum-space states. One might intuitively (though non-rigorously) argue that this quantity could be directly computed for free fermions, and that its value is an adiabatic invariant as interactions are switched on.

Let us determine the change in the symmetry charge of the system in terms  of occupied  (single-particle) fermionic states in the free Fermi gas. First, note that the quasiparticle creation operator $\tilde{c}_{\vecsymb{k}}^\dagger$ transforms under the symmetry $G = \{g|\vecsymb{\tau}\}$ as 
\be
\hat{G} \tilde{c}^\dagger_{\vecsymb{k}} \hat{G}^{-1} = \tilde{c}^\dagger_{g^{-1} \vecsymb{k}} e^{i\vecsymb{\tau}\cdot\vecsymb{k}}
\ee
where $\tilde{n}_{\vecsymb{k}} = \tilde{c}^\dagger_{\vecsymb{k}}\tilde{c}_{\vecsymb{k}}$. Before inserting flux, the system contains a filled Fermi sea: $\ket{\Psi_0} = \prod_{\vecsymb{k}, \text{occ.}} c^\dagger_{\vecsymb{k}} \ket{0}$ and it is easy to show that
\be
\hat{G}\ket{\Psi_0} = \prod_{\vecsymb{k}, \text{occ.}}e^{i \vecsymb{\tau}\cdot \vecsymb{k}}\tilde{c}^\dagger_{g^{-1}\vecsymb{k}}\ket{0}  =  e^{i \sum_{\vecsymb{k}, \text{occ.}} \vecsymb{\tau}\cdot \vecsymb{k} }\ket{\Psi_0},
\ee
where I have used the fact that the set of {all} filled states must be invariant under $\hat{G}$, for otherwise the filled Fermi sea would break symmetry.
Upon inserting the flux, the momentum of each quasiparticle state has shifted as $\vecsymb{k} \rightarrow \vecsymb{k}  + \frac{\vecsymb{K}}{L}$, so that $\ket{\tilde{\Psi}_0} = \prod_{\vecsymb{k}, \text{occ.}} c^\dagger_{\vecsymb{k} +\frac{\vecsymb{K}}{L}} \ket{0}$; therefore, 
\be\label{eq:FLafterflux}
\hat{G}\ket{\tilde{\Psi}_0} =  e^{i \sum_{\vecsymb{k}, \text{occ.}} \left(\vecsymb{\tau}\cdot \vecsymb{k} +\frac{\vecsymb{\tau}\cdot\vecsymb{K}}{L} \right)}\ket{\tilde{\Psi}_0} \equiv e^{i G} \ket{\tilde{\Psi}_0}.
\ee
If one defines $G_0  = \sum_{\vecsymb{k}, \text{occ.}}\vecsymb{\tau}\cdot \vecsymb{k} $, then %we have 
from (\ref{eq:FLafterflux}), %that 
\be\label{eq:FLmomentumcountingapp}
e^{iG} = e^{i\left(G_0 + {N^{(L)}_F}\frac{\vecsymb{\tau}\cdot\vecsymb{K}}{L}\right)}.
\ee
which reduces to the result ~(\ref{eq:FLmomentumcounting})  obtained in the main text.

\section{\label{app:rankcomments}Relation between $\mathcal{S}^*$ and Non-Symmorphic Rank}
As noted in the main text, there is a subtle distinction between the {\it non-symmorphic rank} $\mathcal{S}$ of a space group as defined in Ref.~\cite{Parameswaran:2013ty} and the definition of $\mathcal{S}^*$ used here. While these coincide in most cases, the strict definition of the rank is  the minimum filling $\nu$ at which a symmetric insulating state is possible without triggering fractionalization, and is hence not obviously related to the definition of $\mathcal{S}^*$. In spite of this subtlety, the fact that bands always appear in multiplets of $\mathcal{S}$ in non-symmorphic crystals, suggests that by the heuristic argument on accommodating charge in filled bands versus gapless quasiparticles, $\mathcal{S}$ should replace $\mathcal{S}^*$ in the above expressions.
 However, the precise definition of the rank requires one to consider the electronic polarization, a quantity that is well-defined for an insulator but not in the present case, where there are gapless excitations that carry charge. Therefore, I am % are 
unable at present to give a rigorous justification to replace $\mathcal{S}^*$ by $\mathcal{S}$ in (\ref{eq:mainresult}), and can only conclude that $\mathcal{S}$ must at least be divisible by $\mathcal{S}^*$ and that $\mathcal{S}^*>1$ if and only if the crystal is non-symmorphic. In practice, however, a plurality of non-symmorphic crystals have $\mathcal{S} =2$, and therefore $\mathcal{S}^* = \mathcal{S}$ in these cases.

\end{appendix}
\bibliography{NSLSM_bib}

\end{document}